\documentclass[twocolumn,prl,showpacs]{revtex4}

\usepackage{graphicx}
\usepackage{dcolumn}
\usepackage{float}
\usepackage{amsmath}

\newcommand{\comment}[1]{}

\begin{document}

\title{Evolution of the Spin Susceptibility of High-$T_c$ Superconductors.}

\author{J. Hwang$^{1}$}
\email{hwangjs@mcmaster.ca}
\author{T. Timusk$^{1,2}$}
\author{E. Schachinger$^{3}$}
\author{J. P. Carbotte$^{1,2}$}

\affiliation{$^{1}$Department of Physics and Astronomy, McMaster
University, Hamilton, ON L8S 4M1, Canada\\ $^{2}$The Canadian
Institute of Advanced Research, Toronto, Ontario M5G 1Z8, Canada\\
$^{3}$ Institute of Theoretical and Computational Physics, Graz
University of Technology, A-8010 Graz, Austria}

\date{\today}

\begin{abstract}
We demonstrate that a new tool, a model independent numerical
Eliashberg inversion of the optical self-energy, based on maximum
entropy considerations can be used to extract the magnetic
excitation spectra of high-transition-temperature superconductors.
In Bi$_{2}$Sr$_{2}$CaCu$_{2}$O$_{8+\delta}$ we explicitly show 
that the magnetic mode that dominates the
self-energy at low temperatures directly evolves out of a smooth 
transfer of spectral weight to the mode from the 
continuum just above it.   This redistribution
starts already at 200 K in optimally doped materials but is much 
weaker in overdoped samples. This provides
evidence for the magnetic origin of the superconductivity and
presents a challenge to theories of the spin susceptibility and to
neutron scattering experiments in high-temperature
superconductors.
\end{abstract}


\maketitle

The phenomenon of high-temperature superconductivity, discovered
in the copper oxides twenty years ago~\cite{bednorz86}, continues
to challenge both theorists and experimentalists. From the
beginning it was clear that magnetism was an important part of the
solution to this puzzle.  In conventional superconductors the
spectrum of pairing excitation was successfully extracted as an
electron-phonon spectral function using an inversion of
experimental tunnelling~\cite{carbotte90} and optical
data~\cite{farnworth74} based on the Eliashberg equation. In
principle this spectrum contains the information on the
microscopic interactions among electrons mediated by boson
exchanges which are needed to describe the superconductivity and
can be calculated from band structure
information~\cite{tomlinson76}. The primary tools used to map out
the magnetic excitations in high-temperature superconductors have
been neutron scattering and nuclear magnetic resonance. The
picture that has emerged from these studies is a
not-well-understood spectrum of excitation dominated by a
continuous background extending to unusually high energies,
sometimes called the Millis-Monien-Pines (MMP)
spectrum~\cite{millis90,bourges97,abanov99,fong00,hinkov06} which
evolves into a broad peak in the local (q integrated)
susceptibility at low temperature along with a sharp resonance in
the superconducting state centered at ${\bf q}=(\pi,\pi)$, the 41
meV mode, named after its frequency in optimally doped
YBa$_{2}$Cu$_{3}$O$_{6+\delta}$ (Y-123)
~\cite{rossat-mignod91,mook93,mook98,fong95,dai99,eschrig06}. This
sharp resonance has also been seen in
Tl$_{2}$Ba$_{2}$CuO$_{6+\delta}$~\cite{he02} and in
Bi$_{2}$Sr$_{2}$CaCu$_{2}$O$_{8+\delta}$
(Bi-2212)~\cite{fong99,he01}.  The magnetic
resonance~\cite{norman98,campuzano99,abanov99,zasadzinski06,terashima06}
mode, phonons~\cite{lanzara01,gweon04,zhou05} and the broad
continuum~\cite{millis90,hwang04,carbotte99,schachinger00} in the
bosonic spectra have been proposed as candidates for the pairing
glue in the copper oxides.  Here we use a broader definition of
the magnetic resonance as the peak that develops in the local
magnetic susceptibility at low temperature. Some theories of the
spin susceptibility have predicted that the magnetic resonance
develops out of the MMP continuum~\cite{abanov99,morr98} when the
superconducting gap forms. However, so far no consensus has been
achieved. To throw new light on this issue we have undertaken a
study of the spectrum of excitations responsible for the
self-energy of the carriers as a function of temperature and
doping in three Bi-2212 systems using optical spectroscopy
focussing on the overdoped region of the phase diagram. We clearly
observe experimentally for the first time the evolution of a
strong peak in the integrated magnetic response developing from
the MMP continuum just above it as a function of both {\it temperature} and {\it doping}.

The optical self-energy, $\Sigma^{op}(\omega)$ which involves a
momentum average, is closely related to the quasiparticle
self-energy measured by angle resolved photoemission spectroscopy
which is momentum specific. The optical self-energy is  is defined
in terms of an extended Drude formula~\cite{hwang04}:
\begin{equation}
\sigma(\omega)\equiv
\frac{i}{4\pi}\frac{\omega_{p}^2}{\omega-2\Sigma^{op}(\omega)},
\end{equation}
where $\omega_p$ is the plasma frequency and
$\Sigma^{op}(\omega)\equiv\Sigma_1^{op}(\omega)+i\Sigma_2^{op}(\omega)$.
The optical scattering rate is defined by
$1/\tau^{op}(\omega)\equiv-2\Sigma_2^{op}(\omega)$. The optical
conductivity, $\sigma({\omega})$ can be found through
Kramers-Kronig transformation of the reflectance which is measured
directly. The numerical inversion of the optical scattering rate
is based on an Eliashberg formalism. We start with a deconvolution
of the approximate relation~\cite{schachinger06,dordevic05},
\begin{equation}
\frac{1}{\tau^{op}(\omega;T)}=\int^{\infty}_{0}d\Omega
K(\omega,\Omega;T) I^2\chi(\Omega),
\end{equation}
where $T$ is temperature, $K(\omega, \Omega;T)$ is a kernel
determined from theory, and $I^2\chi(\Omega)$ is the bosonic
spectrum. For the deconvolution we utilized the maximum entropy
method, originated by E. T. Jaynes~\cite{jaynes57}. For the kernel
we use~\cite{schachinger06}:
\begin{equation}
\begin{split}
K(\omega,
\Omega;T)=\frac{\pi}{\omega}\Big[2\omega\coth(\Omega/2T)-(\omega+\Omega)
\\ \times\coth((\omega+\Omega)/2T)+(\omega-\Omega)
\coth((\omega-\Omega)/2T)\Big] \end{split}
\end{equation} for the normal state
and
\begin{equation}
\begin{split}
K(\omega,
\Omega;T=0)=\frac{2\pi}{\omega}\Big\langle(\omega-\Omega)\theta(\omega+2\Delta_0(\vartheta)-\Omega)
\\\times E(\sqrt{1-4\Delta_0^2(\vartheta)/(\omega-\Omega)^2})\Big\rangle_{\vartheta}
\end{split}
\end{equation}
for the superconducting state where
$\langle\cdots\rangle_{\vartheta}$ denotes the angular average
over $\vartheta$ and $E(x)$ is the complete elliptic integral of
the second order. Here
$\Delta_0(\vartheta)=\Delta_0\cos(2\vartheta)$ reflecting the
d-wave symmetry of the superconducting order parameter. For the
superconducting state at finite temperature we adjust the size of
the maximum gap, $\Delta_0$ according to a BCS temperature
variation. After small adjustments using full Eliashberg formalism
and a least squares fit, we obtain the electron-boson spectral
density, $I^2\chi(\omega)$. Details about uniqueness and quality
of the fit are described in the literature~\cite{schachinger06}.

\begin{figure}[t]
  \vspace*{-1.0 cm}%
  \centerline{\includegraphics[width=3.00 in]{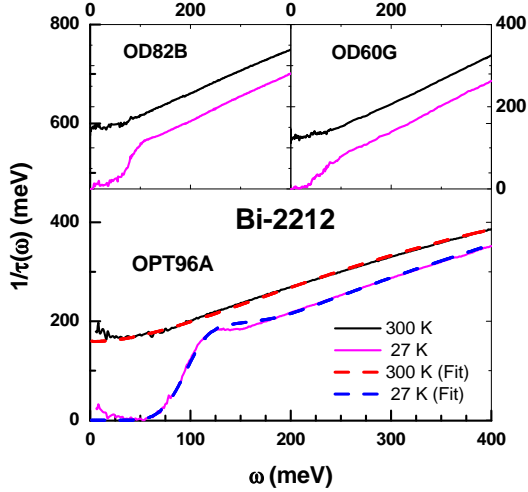}}%
  \vspace*{-2.0 cm}%
\caption{Optical scattering rate, $1/\tau(\omega)$ as a function
of $\omega$. Solid lines are data and dashed lines the Eliashberg
fit. Heavy line is for temperature of 300 K (normal) and light for
27 K (superconducting) for optimally doped Bi-2212. In the inset
we display the optical scattering rate of our two overdoped
samples. They show no low frequency localization effects.}
  \label{Fig1}
\end{figure}

\begin{figure}[t]
  \vspace*{+0.7 cm}%
  \centerline{\includegraphics[width=3.00 in]{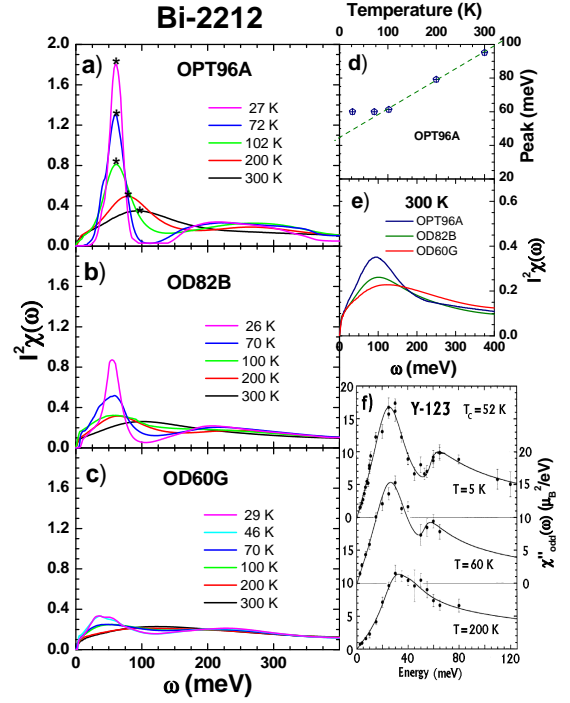}}%
  \vspace*{-0.5 cm}%
\caption{The bosonic spectra, $I^2\chi(\omega)$, of
Bi$_{2}$Sr$_{2}$CaCu$_{2}$O$_{8+{\delta}}$. (a - c) The doping and
temperature dependent bosonic function for three doping levels
(T$_{c}=96$ K (OPT), 82 K (OD), and 60 K (OD)). At room
temperature all samples exhibit a broad continuum background shown
in panel e. On lowering the temperature the broad background peak
evolves into a broad peak with a deep valley above it. The
spectral weight gained in the peak is roughly balanced by the
spectral weight loss in the valley (see Fig.~\ref{Fig3}b). With
increasing doping the intensity of the peak and valley are
significantly weakened (see Fig.~\ref{Fig3}b). Panel d shows the
temperature dependence of the frequency of the peak maximum in
OPT96A marked by a star. Panel f shows the local magnetic
susceptibility of underdoped Y-123 from neutron scattering
~\cite{bourges97,fong00,bourges99}. While this is a different
copper oxide from the ones given in panels a, b and c we note the
qualitative similar temperature evolution of panel a.}
  \label{Fig2}
\end{figure}

We used the self-energy data of Bi-2212 from a previous study~\cite{hwang04}
on an optimally doped and two overdoped samples. We note that the
optimally doped sample is yttrium-doped and somewhat different
from conventional optimally doped Bi-2212 having a more ordered
structure and a higher $T_c$. Underdoped systems show pseudogap
behavior not yet incorporated in the inversion formalism and are
not treated here. In Fig.~\ref{Fig1} we show our fit (dashed
lines) to data (solid lines) as an example for our optimally doped
sample OPT96A at two temperatures: heavy lines are for $T$=300 K
(normal) and light lines for $T$=27 K (superconducting). The fits
are very good except for two small spectral regions near zero
energy for both states and near the overshoot region for the
superconducting state. Above 125 meV in the superconducting state
the theoretical curve is flat while the data show a slight
depression. The discrepancy near zero energy may come from
impurity localization which is observed in most cuprate systems.
We note that the optimally doped sample shows the strongest
localization but this does not affect to the main sharp peak in
$I^2\chi(\Omega)$. The two overdoped samples show negligible
localization (see the insets in Fig. 1).

In Fig.~\ref{Fig2}a, b and c we plot the extracted bosonic
spectra, $I^2 \chi(\omega)$, of Bi-2212. We observe dramatic
temperature and doping dependencies. The room temperature spectra
show a broad continuum which exhibits some doping dependence as
shown in panel e. At lower doping levels the broad peak in the
spectrum becomes somewhat stronger and shifts to lower
frequencies, which is consistent with a previous neutron
study~\cite{fong00} on underdoped YBa$_2$Cu$_3$O$_{6.6}$. The
temperature and doping dependence, as we move from panel a to
panel c, is more striking. On the lowering of the temperature
nearly all the spectral weight below 130 meV in the optimally
doped material is moved to a resonance peak while in the overdoped
region less and less of the continuum spectral weight is
transferred to this peak. The peak in the room temperature
spectrum, at 97 meV in panel a, shifts continuously to lower
energies as its intensity grows. The panel d shows that the peak
position is approximately linear in temperature above $T=100$ K
and saturates to 60 meV below this temperature. What is notable
about this process is that this evolution takes place through a
transfer of spectral weight from frequencies above the peak
leaving a valley between 100 to 150 meV. In addition, there is a
well developed 30 meV gap on the low frequency side of the peak.
Such a gap in the spin susceptibility is consistent with the
observations of spin-polarized neutron scattering of optimally
doped Y-123 with $T_c$=91 K~\cite{bourges99}. As shown in
Fig.~\ref{Fig2}b and ~\ref{Fig2}c for higher doping levels the gap
in the spin susceptibility becomes small and the resonance mode
can still be discerned although it weakens
significantly~\cite{hwang04}. In our most overdoped sample (OD60G)
only a small fraction of spectral weight in the bosonic spectrum
contributes to the temperature redistribution.

\begin{figure}[t]
  \vspace*{-0.5 cm}%
  \centerline{\includegraphics[width=3.00 in]{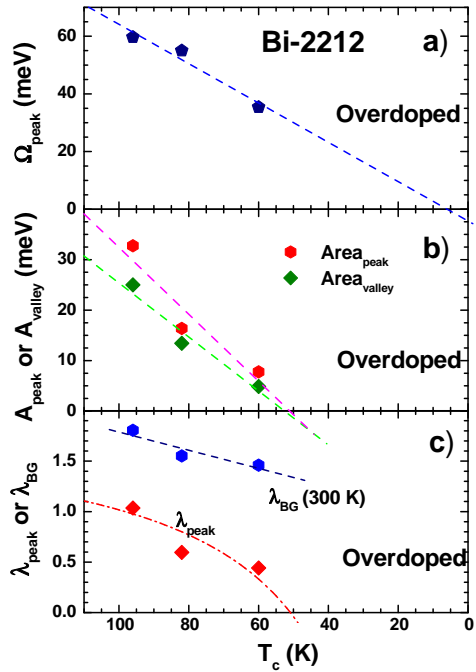}}%
  \vspace*{-0.5 cm}%
\caption{$T_c$ dependent properties of the peak and the valley
(see in the text). a) The central peak frequency is proportional
to $T_c$, i.e., $\Omega_{peak}=8.0 k_B T_c$, and the dashed line
is a linearly fitted line. b) The peak and valley are closely
connected, have a very similar $T_c$ dependence and vanish at the
same $T_c\approx 50$ K. The dashed lines are linearly fitted
lines. c) The coupling constant, $\lambda_{peak}$, vanishes at the
same $T_c$ as the peak and valley. The coupling constant,
$\lambda_{BG}$, of the continuous background shows a weaker $T_c$
dependence as compared to that of the peak. The dash-dotted line
is obtained by using the linearly fitted center frequency
(Fig.~\ref{Fig3}a) and area under the peak (Fig.~\ref{Fig3}b),
i.e., $\lambda_{peak}\simeq 2 A_{peak}/\Omega_{peak}$.}
 \label{Fig3}
\end{figure}

We also show the local (q integrated) magnetic susceptibility,
$\chi_{odd}''(\omega)$, from neutron scattering study of
underdoped Y-123~\cite{bourges97,fong00} in Fig.~\ref{Fig2}f to
compare its temperature dependent spectral weight redistribution
with that of the bosonic excitation, $I^2 \chi(\omega)$, of our
optimally doped Bi-2212 (OPT96A). Although they are different
copper oxide systems and at different doping levels they
qualitatively contain common temperature dependent features which
we described previously. A similar spectral redistribution of the
local magnetic susceptibility has been observed by Dai {\it et
al.}~\cite{dai99}. However, it was not widely recognized or
confirmed by other experiments. Here we find that the spectral
redistribution occurs mainly in the low frequency region below 200
meV and is strongly doping dependent on the overdoped side of the
phase diagram.

Alternatively, we can look at the spectrum at low temperature in
panel a of Fig.~\ref{Fig2} as the development of a large 130 meV
gap in the spin fluctuation spectrum with a mode in the middle of
the gap at 60 meV. Such a transformation has been considered by
Abanov and Chubukov~\cite{abanov99} but with the gap arising at
2$\Delta_0$ with $\Delta_0$ the superconducting gap. Here we find
a much larger gap that starts to form already at $T=200$ K, well
above $T_c$.

To study quantitatively the changes in the bosonic spectra we
subtract the room temperature spectrum from the lower temperature
spectra. We clearly observe a peak and a valley in the difference
spectra. In Fig.~\ref{Fig3} we display the $T_c$ dependence of the
peak and valley in the bosonic spectra. Fig.~\ref{Fig3}a shows the
center frequency of the  peak, $\Omega_{peak}$ versus $T_c$ at our
lowest measured temperatures. We find $\Omega_{peak} \approx 8.0
k_BT_c$, which is similar to the temperature dependence of the
magnetic resonance mode ($\Omega_{res}\approx 5.3k_BT_c$). In
Fig.~\ref{Fig3}b we plot the the area under the peak and of the
valley above it as a function of the critical temperature for our
lowest temperature (for the optimally doped we also include the 30
meV gap as part of the valley). The two areas show the same trend;
both decrease rapidly as $T_c$ decreases in the overdoped region
and their extrapolation show that they vanish at the same $T_c$
near 50 K. The total spectral weight up to 400 meV is conserved to
within 15 \% which indicates a small frequency dependence of the
coupling constant $I^2$ in $I^2\chi(\Omega)$. We also calculate
the $T_c$ dependent coupling constant,
$\lambda(T_c)=2\int^{\omega_c}_{0}I^2\chi(\Omega)/\Omega d\Omega$,
using only the peak ($\lambda_{peak}$) or the continuum
($\lambda_{BG}$) for $I^2\chi(\Omega)$. Here $\omega_c$ is a
cutoff frequency, taken as 400 meV. These are shown in
Fig.~\ref{Fig3}c. The peak coupling constant $\lambda_{peak}$
decreases rapidly extrapolating to zero at $T_c$ =50 K. The
coupling constant of the background $\lambda_{BG}$ shows a weaker
dependence on $T_c$.

Our novel analysis of the optical data has provided new and
detailed information on the redistribution with temperature and
doping levels of the spectral weight in the inelastic scattering
spectral function. This analysis of optical data, based on an
Eliashberg inversion, is the only existing method to date which
can be used to study the evolution of the spin susceptibility as a
function of doping and temperature across the phase diagram of
high-$T_c$ oxides. Furthermore this
analysis can also be applied to other existing optical spectra to
get new information on their excitation spectrum. Our central new
finding is that the optical resonance peak which becomes prominent
at low temperature in our optimally doped sample, forms mainly
through the transfer of spectral weight from the energy region
immediately above it. This constitute strong evidence that the
peak and background in the bosonic spectral function have the same
microscopic origin. Identifying the optical peak with the
resonance in the local magnetic susceptibility measured by neutron
scattering, we conclude that the background which dominates the
inelastic interaction at and above $T_c$ is magnetic in origin.

Finally we note that we have also identified a new temperature
scale, around 200 K in the optimally doped Bi-2212, where the
magnetic fluctuation spectrum begins to soften and develop the
peak in the local magnetic susceptibility. To our knowledge
current theories do not account for this phenomenon.

\acknowledgments This work has been supported by the Canadian
Natural Science and Engineering Research Council and the Canadian
Institute of Advanced Research. We acknowledge the contribution of
Genda Gu and Hiroshi Eisaki in growing the crystals that we used.
We have profited from discussions with Philip Anderson, Philip
Bourges, Andrey Chubukov, Matthias Eschrig, Berhnard Keimer and
Chandra Varma.

\bibliography{scibib}

\end{document}